\documentclass[%
 reprint,
superscriptaddress,
 amsmath,amssymb,
 aps,
showkeys
]{revtex4-2}

\usepackage{graphicx}
\usepackage{dcolumn}
\usepackage{bm}

\usepackage{xcolor}
\usepackage{soul}

\usepackage{multirow}
\usepackage{subfigure}

\begin{document}

\preprint{APS/123-QED}

\title{Electronic properties of substitutional impurities in graphene-like C$_2$N, $tg$-C$_3$N$_4$, and $hg$-C$_3$N$_4$}

\author{Saif Ullah}
 \email{sullah@fisica.ufjf.br}
 \affiliation{Departamento de F\'isica, Instituto de Ci\^encias Exatas, Campus Universit\'ario, Universidade Federal de Juiz de Fora, 36036-900 Juiz de Fora, MG, Brazil}
 
\author{Pablo A. Denis}
 \affiliation{Computational Nanotechnology, DETEMA, Facultad de Qu\'imica, UDELAR,
CC 1157, 11800 Montevideo, Uruguay}

\author{Marcos G. Menezes}
 \email{marcosgm@if.ufrj.br}
 \affiliation{Instituto de F\'isica, Universidade Federal do Rio de Janeiro, Caixa Postal 68528, 21941-972 Rio de Janeiro, RJ, Brazil}
 
\author{Fernando Sato}
 \affiliation{Departamento de F\'isica, Instituto de Ci\^encias Exatas, Campus Universit\'ario, Universidade Federal de Juiz de Fora, 36036-900 Juiz de Fora, MG, Brazil}
 
\author{Rodrigo B. Capaz}
 \affiliation{Instituto de F\'isica, Universidade Federal do Rio de Janeiro, Caixa Postal 68528, 21941-972 Rio de Janeiro, RJ, Brazil}

\date{\today}

\begin{abstract}
We study the electronic and structural properties of substitutional impurities of graphene-like nanoporous materials C$_2$N, $tg$-, and $hg$-C$_3$N$_4$ by means of density functional theory calculations. We consider four types of impurities; boron substitution on carbon sites (B(C)), carbon substitution on nitrogen sites (C(N)), nitrogen substitution on carbon sites (N(C)), and sulfur substitution on nitrogen sites (S(N)). From cohesive energy calculations, we find that the C(N) and B(C) substitutions are the most energetically favorable and induce small bond modifications in the vicinity of the impurity, while the S(N) induces strong lattice distortions. Though all of the studied impurities induce defect levels inside the band gap of these materials, their electronic properties are poles apart depending on the behavior of the impurity as an acceptor (B(C) and C(N)) or a donor (N(C) and S(N)). It is also observed that acceptor (donor) wavefunctions are composed only of $\sigma$ ($\pi$) orbitals from the impurity itself and/or neighboring sites, closely following the orbital composition of the valence (conduction) band wavefunctions of the pure materials. Consequently, acceptor wavefunctions are directed towards the pores and donor wavefunctions are more extended throughout the neighboring atoms, a property that could further be explored to modify the interaction between these materials and adsorbates. Moreover, impurity properties display a strong site sensitivity and ground state binding energies ranging from $0.03$ to $1.13$ eV in non-magnetic calculations, thus offering an interesting route for tuning the optical properties of these materials. Finally, spin-polarized calculations reveal that all impurity configurations have a magnetic ground state with a total moment of $1.0 \ \mu_B$ per unit supercell, which rises from the spin splitting of the impurity levels. In a few configurations, more than one impurity level can be found inside the gap and two of them could potentially be explored as two-level systems for single-photon emission, following similar proposals recently made on defect complexes on TMDCs.
\end{abstract}


\maketitle


\section{\label{sec:intro}Introduction}

The experimental realization of the first-ever 2D material, graphene, took more than half a century since it was first studied theoretically \cite{novoselov2004electric,wallace1947band}. Now, within a timeframe of less than a couple of decades, the existence of several hundreds of other 2D materials have already been reported or theoretically proposed \cite{mounet2018two}. This rapidly growing family of materials now includes hexagonal boron nitride ($h$-BN) \cite{watanabe2004direct,lin2010soluble}, silicene \cite{vogt2012silicene}, phosphorene \cite{liu2014phosphorene}, borophene \cite{mannix2015synthesis}, MXenes \cite{naguib2012two, khazaei-mxenes-2019}, antimonene \cite{zhang2015atomically}, 
 and many others. Their electronic and optical properties vary broadly, which enables them to be utilized in a broad spectrum of applications.
 

Among them, graphene is naturally the most well-known material and offers several interesting properties with potential applications in numerous fields \cite{novoselov2004electric, novoselov2007room, neto2009electronic,lin2010100, cheng2012high}. However, graphene cannot be used in FETs due to the absence of an energy gap \cite{geim2010rise}. This deficiency can be remedied, for instance by doping the sheet with substitutional atoms or adsorbates. In particular, nitrogen (N) doping is a very interesting alternative due to the comparable radii of C and N and the fact that it allows the redesign of many properties of graphene nanosheets and nanostructures \cite{luo2011pyridinic, wei2009synthesis, wang2010nitrogen, lv2012nitrogen, zhang2013dimension, gong2009nitrogen, ullah2018first, ullah2019adsorption, ullah2017triple}. Besides, there is an additional benefit as N can be integrated at multiple positions in graphene, leading to different functionalities and a wide variety of structures of strongly bonded organic frameworks \cite{luo2011pyridinic, yasuda2013selective, xu2016synergistic}. Consequently, there exists a broad array of synthesized carbon nitride structures.

Recently, the synthesis of carbon nitride monolayers of various stoichiometries, including C$_2$N \cite{mahmood2015nitrogenated}, C$_3$N \cite{mahmood2016two}, and C$_3$N$_4$ \cite{thomas2008graphitic, algara2014triazine}, have been reported. These novel 2D materials display many interesting electronic, optical, thermal, mechanical, and magnetic properties \cite{li2017directional, cui2017steering}. In particular, C$_2$N can be regarded as a nitrogenated nanoporous or holey graphene, in which structural holes or vacancies are introduced in the original honeycomb structure, leading to a 2:1 proportion of C:N and an equal sharing of N and vacancy sites \cite{mahmood2015nitrogenated, zhang2015effects, xu2015two}, as shown in Fig. \ref{fig:pure}. This makes the adsorption of atoms, ions, and even larger molecules favorable. Consequently, C$_2$N can act as a high selective nano-filter \cite{mahmood2015nitrogenated, mortazavi2016thermal}. Moreover, C$_3$N$_4$ can be found in a number of different phases including cubic and semi-cubic phases, alpha and beta phases, and graphitic phases \cite{martha2013facile, dong2017tunable, mortazavi2015mechanical}. The latter are divided into further two classes, $tg$-C$_3$N$_4$ and $hg$-C$_3$N$_4$, and are considered as the most stable phases with semiconducting properties. Their structures are also shown in Fig.  \ref{fig:pure}. These materials possess excellent electronic, optical, and sensing properties and are strong photo-catalysts for water splitting \cite{zhu2014lithium, li2013stability}.

Despite the interesting electronic properties and desirable applications of these materials, doping effects have not been investigated so far. As a matter of fact, to our knowledge, there are no reports in the literature on the electronic properties of donor and acceptor impurities in these systems \cite{bafekry2020first, makaremi2018band}. Since doping can be used for tailoring the electronic and optical properties of semiconductors, it is crucial to developing an in-depth understanding of dopants and their properties. Specifically, we must look at the impurity levels induced by these dopants inside the band gap, as their presence may give rise to new optical transitions that are very important for applications in optoelectronic devices. For example, in $h$-BN, a C impurity replacing a N atom can redesign its optical gap by allowing new transitions between the impurity level and the conduction band, resulting in an extra peak at about $4$ eV \cite{katzir-prb,moore-jphyschem, museur-prb, bourrellier-nanolett}. Additionally, sometimes these impurity levels even come with exotic physics, such as the sublattice sensitivity found in $h$-BN \cite{ullah2019exotic}. Other types of impurities were also explored in h-BN, both at theoretical and experimental levels, and are also found to induce important modifications in the electronic and optical properties of this material. Examples include single vacancies, vacancy-substitutional defect complexes, and interstitial impurities \cite{weston-prb-2018, mehdi-acsphoto-2018, jiandong-nanolett-2018, qixing-nanolett-2018}.

In this work, we employ first-principles calculations to extensively study the electronic and structural properties of substitutional dopants in graphene-like nanoporous materials C$_2$N, tg-, and hg-C$_3$N$_4$. We consider four types of dopants: B, C, N, and S, two of which act as acceptors and the other two as donors. As we shall see, the electronic properties of the acceptor and donor levels found inside the gap are at odds and are closely related to the contrasting orbital compositions of the valence and conduction band of these materials. Additionally, a strong site sensitivity is also found, in a similar fashion to $h$-BN, with binding energies that ranges from $0.03$ to $1.13$ eV. Therefore, substitutional doping provides a compelling route for tuning the electronic and optical properties of these materials. Finally, magnetic configurations and the spin splittings of the impurity levels were also studied. 

The article is organized as follows. In the next section, we present the methods of our calculations based on density functional theory. In section \ref{sec:results}, we present and discuss our results, by considering separately various important aspects such as the electronic properties of the pure systems, structural properties and stability of dopants, electronic properties of the impurity levels in non-magnetic configurations and, finally, the effects of spin-polarization. Our conclusions are presented in section \ref{sec:conclusions}.

\section{\label{sec:method} Methods}

We perform first-principles  calculations based on density functional theory (DFT) \cite{ho-kohn, kohn-sham} to study the electronic and structural properties of substitutional impurities in C$_2$N, $tg$-C$_3$N$_4$, and $hg$-C$_3$N$_4$. The SIESTA code is employed in the calculations unless otherwise stated \cite{siesta, ordejon1996self}. We use the vdW-DF1 \cite{dion-vdw-DF, perez-vdw-DF} exchange-correlation functional which, despite the absence of relevant van-der-Waals interactions in single-layer systems, has a better performance in describing band gaps than the local (LDA) and semi-local (PBE-GGA) exchange-correlation functionals, as a consequence of its fully non-local nature \cite{wang1983density, chan2010efficient}. Norm-conserving (NC) Troullier-Martins (TM) \cite{troullier-martins} pseudopotentials in fully-relativistic form were used to treat the ion-electron interactions. We include a large vacuum of more than $20$ \AA \ in the out-of-plane direction of our cells in order to eliminate pseudo-interactions between periodic images. The Brillouin zones (BZ) of pristine C$_2$N, $tg$-C$_3$N$_4$, and $hg$-C$_3$N$_4$ were sampled with $8 \times 8 \times 1$, $15 \times 15 \times 1$ and $9 \times 9 \times 1$ Monkhorst-Pack k-point grids, respectively, and denser grids were used for the density of states calculations (DOS and PDOS) \cite{monkhorst-pack}. For the impurity calculations, we employ supercells as explained in detail in section \ref{subsec:stability}. In order to sample the BZ of these supercells, we choose a Monkhorst-Pack k-point grid of $3 \times 3\times 1$ in all cases, as the supercell sizes of the three structures are quite similar. For density of states calculations, we use a denser k-point grid of $11 \times 11 \times 1$ and a broadening of $0.01$ eV. We use a mesh cutoff of $200$ Ry for real space projections accompanied by a double zeta polarization (DZP) basis set. All the structures were first relaxed in spin-polarized calculations until all forces acting on the atoms were smaller than $0.02$ eV/\AA. The energy convergence criterion is selected as $10^{-6}$ eV. In a second step, non-magnetic configurations were further relaxed taking the optimized configurations from the polarized calculations as a starting point. As such, all structures considered for electronic structure calculations in this work are fully optimized in both magnetic and non-magnetic scenarios. Geometry optimizations were performed with the conjugate gradient (CG) scheme.
Finally, some of the results for selected defect configurations were cross-checked against auxiliary calculations with a PBE-GGA functional and PAW pseudopotentials in the VASP code, for which a very good agreement is found \cite{kresse1994ab,perdew1996generalized,blochl1994projector,kresse1996efficiency}. These results are shown in the supplementary material file (SM).

\section {\label{sec:results} Results}

\subsection{\label{subsec:pure} Electronic and Structural Properties of Pristine Materials}

\begin{figure*}
    \centering
    \includegraphics[width=\textwidth]{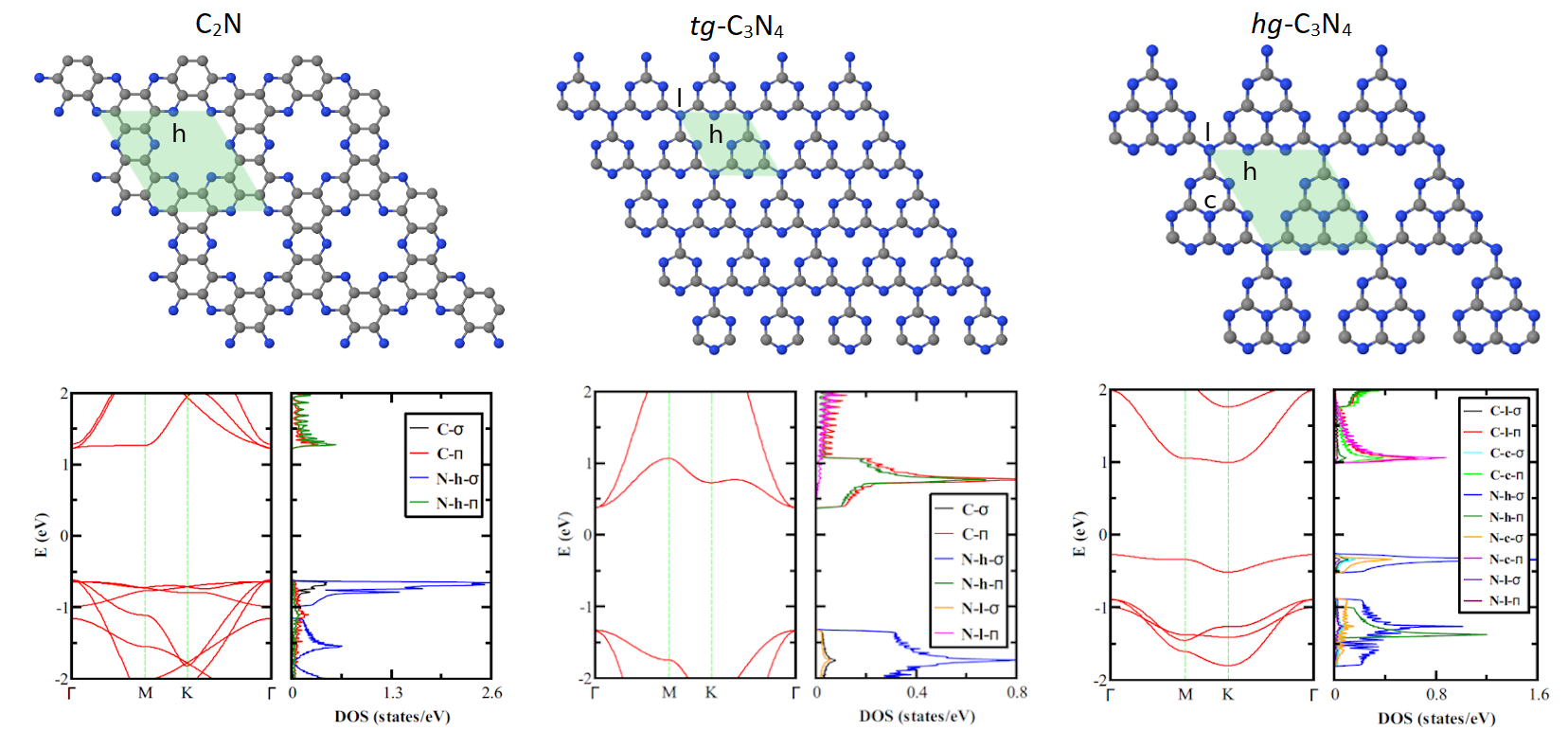}
    \caption{\label{fig:pure} Top: crystal structures of C$_2$N (left), $tg$-C$_3$N$_4$ (middle) and $hg$-C$_3$N$_4$ (right). The green areas represent the unit cells and the non-equivalent nitrogen atoms are labelled as $h$ (hole), $l$ (link) and $c$ (center) in each structure. The $hg$-C$_3$N$_4$ structure has two non-equivalent carbon atoms, labeled according to the neighboring N$_l$ or N$_c$ sites. Bottom: band structures and projected density of states (PDOS) for selected atoms in the corresponding structures. The contributions from $\sigma$ and $\pi$ orbitals from all non-equivalent atoms are included, as specified in the color codes.}
\end{figure*}

We begin by discussing the pristine C$_2$N, $tg$-C$_3$N$_4$ and $hg$-C$_3$N$_4$ structures. Their crystal structures are shown in Fig. \ref{fig:pure} and the corresponding lattice constants and bond lengths are listed in Table \ref{tab:pure}. The unit cell of C$_2$N consists of 12 carbon (C) and 6 nitrogen (N) atoms. All N atoms are equivalent to each other and lie at the edges of the pores (or holes) and, therefore, we label these atomic sites as hole ($h$) sites. Likewise, all C atoms have the same chemical environment and are equivalent to each other. The hole size (diameter) can be measured by the distance between opposite N atoms, which is $5.5$ \AA. This value is a bit smaller than the separation between oxygen (O) atoms in nanoporous boroxine (B$_3$O$_3$) \cite{ullah2019theoretical-B3O3}.

The unit cell of $tg$-C$_3$N$_4$ consists of 3 C and 4 N atoms. As we can see in Fig. \ref{fig:pure}, this structure has two non-equivalent N sites, but all C atoms are equivalent. 3 N atoms in the unit cell lie at the hole edges and are, again, labeled as hole ($h$) sites, while the fourth N atom bonds with 3 C atoms and is labeled as a link ($l$) site. Therefore, the structure has two different C-N bond lengths, as reported in Table \ref{tab:pure}.

Finally, the unit cell of $hg$-C$_3$N$_4$ consists of 6 C and 8 N atoms. It has the same shape as the $tg$-C$_3$N$_4$ structure, but the triangular motifs are larger and contain a third non-equivalent N atom at their centers, as shown in Fig. \ref{fig:pure}. For that reason, we label this site as a center ($c$) site. Each unit cell has 6 N atoms in $h$ sites, 1 in a $l$ site and 1 in a $c$ site. Moreover, note that this structure has two non-equivalent C sites. Half of the C atoms makes one bond with a N atom at a $l$ site, while the other half makes one bond with a N atom at a $c$ site. For this reason, we also label these C atomic sites as $l$ and $c$, respectively. The structure has three different C-N bond lengths, as reported in Table \ref{tab:pure}. The geometrical properties of all structures are in good agreement with previous experimental and theoretical studies \cite{mahmood2015nitrogenated, zhu2015c, zhang2015effects, qu2016highly, xu2015two, bafekry2020first, makaremi2018band, thomas2008graphitic, xu2012band} and with our cross-check calculations reported in the SM file.

\begin{table}[h]
    \setlength{\tabcolsep}{4.0pt}
    \setlength{\extrarowheight}{3.0pt}
    \centering
    \begin{tabular}{|l|c|c|c|c|}
        \hline
        Material      & $a$ (\AA) & $d_{C-C}$ (\AA)   & $d_{C-N}$ (\AA)                          & $E_g$ (eV) \\
        \hline
        C$_2$N        & $8.35$    & $1.43$, $1.48$    & $1.35$ ($h$)                             & $1.84$     \\
        $tg$-C$_3$N$_4$ & $4.82$    &  -                & $1.34$ ($h$), $1.47$ ($l$)               & $1.71$     \\
        \multirow{2}{*}{$hg$-C$_3$N$_4$} & \multirow{2}{*}{$7.18$} & \multirow{2}{*}{-}            & $1.35$ ($h$), $1.49$ ($l$) & \multirow{2}{*}{$1.27^*$}   \\
                                                                                             & & & $1.40$ ($c$) &   \\
        \hline
    \end{tabular}
    \caption{\label{tab:pure} Lattice constants ($a$), bond lengths ($d_{C-C}$ and $d_{C-N}$) and band gaps ($E_g$) of the pure structures. The star next to the value indicates an indirect gap. For a description of the non-equivalent nitrogen sites, see the text and Fig. \ref{fig:pure}.}  
\end{table}

Now we move on to the electronic properties of these materials, which are also shown in Fig. \ref{fig:pure}. We can see that all of them are semiconductors, with Kohn-Sham band gaps between $1.27$ and $1.84$ eV, as reported in Table \ref{tab:pure}. The $hg$-C$_3$N$_4$ structure has an indirect band gap, with the valence band maximum (VBM) and conduction band minimum (CBM) at the $\Gamma$ and $K$ points of the Brillouin Zone (BZ), respectively. On the other hand, the other two structures have direct band gaps at the $\Gamma$ point. 
These properties are in good agreement with previous calculations at PBE, LDA, and GW levels \cite {xu2012band, bafekry2020first, makaremi2018band}. Naturally, the absolute value of the gaps depends on the choice of method, but the semiconducting behavior and the direct/indirect nature of the gaps agree with our calculations.

The orbital analysis is done by calculating the projected density of states (PDOS) calculations, as shown in Fig. \ref{fig:pure}. The results are quite interesting, as they show that the orbital compositions of the valence and conduction bands are very different. In all cases, the valence bands are composed solely of $\sigma$ orbitals, that is, in-plane $sp^2$ orbitals. Moreover, the most important contribution comes from N atoms at hole ($h$) sites, which have unpaired, localized $\sigma$ electrons. This explains the low dispersion of these bands in the C$_2$N and $hg$-C$_3$N$_4$ structures, which have bigger holes and a larger number of N$_h$ sites. In the $tg$-C$_3$N$_4$ structures, these sites are closer and the unpaired orbitals have a stronger overlap, leading to a larger bandwidth.

In contrast, the conduction bands of these materials are composed only of out-of-plane $\pi$ orbitals. In the C$_2$N and $tg$-C$_3$N$_4$ structures, the C and N$_h$ sites give the most important contributions in nearly equal proportions, while in the $hg$-C$_3$N$_4$, the most important contributions come from C$_c$, C$_l$ and N$_c$ sites. As these sites are close to each other, the overlaps are stronger and the bands are more dispersive. As we shall see, these striking differences between the valence and conduction bands will be central to our discussion of the electronic properties of donor and acceptor impurities in these materials.

\subsection{\label{subsec:stability} Stability and Structural Properties of Substitutional Impurities}

In order to study the electronic and structural properties of substitutional impurities in these materials, we use the supercell approximation. We place a single impurity in a $3 \times 3$, $5 \times 5$ or $3 \times 3$ supercell of C$_2$N, $tg$-C$_3$N$_4$ or $hg$-C$_3$N$_4$, respectively. These cells contain $162$, $175$ and $126$ atoms, respectively. We replace a C atom in a supercell with a boron (B) atom, which behaves as an acceptor, or a nitrogen (N) atom, which behaves as a donor. Likewise, we replace a N atom with a carbon (C) or a sulfur (S) atom. For a given structure, we label each impurity case as X(Y), where X is the chemical species of the impurity and Y is the replaced atom so, for instance, C(N$_h$) represents a C impurity replacing a N atom at a hole site. We consider all the non-equivalent sites in each structure, following the notation introduced in the previous section.

We show the cohesive energies and the modified bond lengths between the impurity and its first neighbors in Table \ref{tab:energetics}. The cohesive energy is defined as:

\begin{equation}
    E_c = \dfrac{E_{t} - \sum_i n_i E_i}{N},
\end{equation}

\noindent where $E_{t}$ is the total energy of the supercell with (or without) the impurity. The number of atom(s) and the corresponding energies of species $i$ {color{green} (in isolated form) } are represented by $n_i$ and $E_i$, respectively. Finally, $N$ represents the total number of atoms inside the supercell. 

%



From an experimental point of view, another quantity of interest is the formation energy, defined as:


\begin{equation}
   E_f = \dfrac{E_{t}- \sum_i n_i \mu_i}{N}
\end{equation}

\noindent where $\mu$ represents the chemical potential of species $i$, which may be an atom from the original structure or the impurity. The chemical potentials of B, C, N, and S are taken as the energy of a single atom in the boron alpha-sheet, graphene, N2 gas, and bulk-sulfur, respectively. Other definitions of formation energy may consider different chemical potentials, so care must be taken when making comparisons. With our definition, the energetics given by SIESTA, as reported in Table \ref{tab:energetics}, match quite well with our cross-check calculations in VASP, which can be found in the SM. As we can see, their values range from 0.21 to 0.32 eV and display both a material sensitivity and a weak defect-type sensitivity, in a similar fashion to the behavior of the cohesive energies, which we discuss in more detail below.
  



From the point of view of cohesive energy, we can see that the C(N) substitution, in all three structures,  has a greater strength. In fact, the cohesive energies found in these cases are even smaller than those found in the pure structures. When the C atom replaces a N$_h$ (N$_l$) atom, the C-C bonds between the impurity and its first neighbors are slightly expanded (contracted) in comparison with the C-N bonds found in the pure structures and are brought closer to the length found in graphene. For the C(N$_c$) substitution in the $hg$-C$_3$N$_4$ structure, the bond length remains unchanged. The B(C) substitution is also quite robust, as the cohesive energies are smaller than the pure values for the $tg$-, and $hg$-C$_3$N$_4$ structures and slightly higher for the C$_2$N structure. The bonds in the vicinity of this impurity are all expanded in comparison with those in the pure structures and the B-N$_l$ and B-N$_c$ bond lengths are close to the value found in h-BN.

In contrast, N(C) and S(N) show the lowest cohesive strength in comparison with the other substitutions in all structures. Like C(N) and B(C), the N(C) substitution induces small bond length modifications in the vicinity of the impurity, while the S(N) impurity induces large distortions as a consequence of the large radius of the S atom. Nevertheless, the relative difference in cohesive strength is quite small which shows the excellent stability and robustness of these structures.

Considering the ease of experimental synthesis as given by the formation energies, we can see that the B(C) substitution is the most favorable in C$_2$N, closely followed by S(N) and N(C). The least favorable case is the C substitution on N site. Furthermore, the B(C) substitution bears the lowest formation energy in the C$_3$N$_4$ structures as well. Notice, however, that the formation energy profile in these structures is different than the one found in C$_2$N. For these cases, the second most favorable substitution is C(N), while the S(N) and N(C) substitutions are less favorable.
\begin{table*}[!htbp]
    \setlength{\tabcolsep}{4.5pt}
    \setlength{\extrarowheight}{3.0pt}
    \centering
    \begin{tabular}{|l|c|c|c|c|}
        \hline
        Material                & Impurity & $E_c$ (eV/at)   & $E_f$ (eV/at) & $d_{1st}$ (\AA)  \\
        \hline
        \multirow{6}{*}{C$_2$N} & pure     & -6.409       & -              & - \\ \cline{2-5}
                                & \multirow{2}{*}{B(C)}   & \multirow{2}{*}{-6.397} &\multirow{2}{*}{0.207}    & $1.39$ (B-N) \\
                                &                         &                &                                   & $1.57$, $1.54$ (B-C) \\
                                & C(N)     & -6.413       & 0.218          & $1.38$                   \\ \cline{2-5}
                                & N(C)     & -6.385       & 0.213          & $1.40$ (N-N), 1.42 (N-C) \\
                                & S(N)     & -6.394       & 0.211          & $1.75$                 \\
        \hline
        \multirow{7}{*}{$tg$-C$_3$N$_4$} & pure   & -5.675       & -       & - \\ \cline{2-5}
                                       & B(C)     & -5.679       & 0.283   &  $1.42$ ($h$), $1.49$ ($l$) \\
                                       & C(N$_h$) & -5.681       & 0.306   & $1.39$                   \\
                                       & C(N$_l$) & -5.686       & 0.301   & $1.46$                   \\ \cline{2-5}
                                       & N(C)     & -5.639       & 0.317   & $1.37$ ($h$), $1.47$ ($l$) \\
                                       & S(N$_h$) & -5.648       & 0.314   & $1.74$                   \\
                                       & S(N$_l$) & -5.638       & 0.324   & $1.76$ (average)         \\
        \hline
        \multirow{11}{*}{$hg$-C$_3$N$_4$} & pure   & -5.711       & -        & - \\ \cline{2-5}
                                        & B(C$_l$) & -5.716       & 0.242    & $1.42$ ($h$), $1.51$ ($l$) \\
                                        & B(C$_c$) & -5.714       & 0.244    & $1.42$ ($h$), $1.46$ ($c$) \\
                                        & C(N$_h$) & -5.717       & 0.276    & $1.40$                   \\
                                        & C(N$_l$) & -5.728       & 0.265    & $1.46$                   \\
                                        & C(N$_c$) & -5.727       & 0.266    & $1.41$                   \\ \cline{2-5}
                                        & N(C$_l$) & -5.663       & 0.287    & $1.37$ ($h$), $1.48$ ($l$) \\
                                        & N(C$_c$) & -5.664       & 0.286    & $1.34$ ($h$), $1.44$ ($c$) \\
                                        & S(N$_h$) & -5.687       & 0.272    & $1.79$ (average)         \\
                                        & S(N$_l$) & -5.669       & 0.289    & $1.81$                   \\
                                        & S(N$_c$) & -5.647       & 0.312    & $1.71$                   \\
        \hline
    \end{tabular}
    \caption{\label{tab:energetics} Cohesive ($E_c$) and formation ($E_f$) energies for each impurity case and modified bond lengths between the impurity and its first neighbors ($d_{1st}$). The cohesive energies of the pure systems in the corresponding supercells are also included. See the text and Fig. \ref{fig:pure} for a description of the non-equivalent sites.}
\end{table*}

\subsection{\label{subsec:electronics} Electronic Properties of Substitutional Impurities}

\begin{figure}
\centering
    \includegraphics[width=0.5\textwidth]{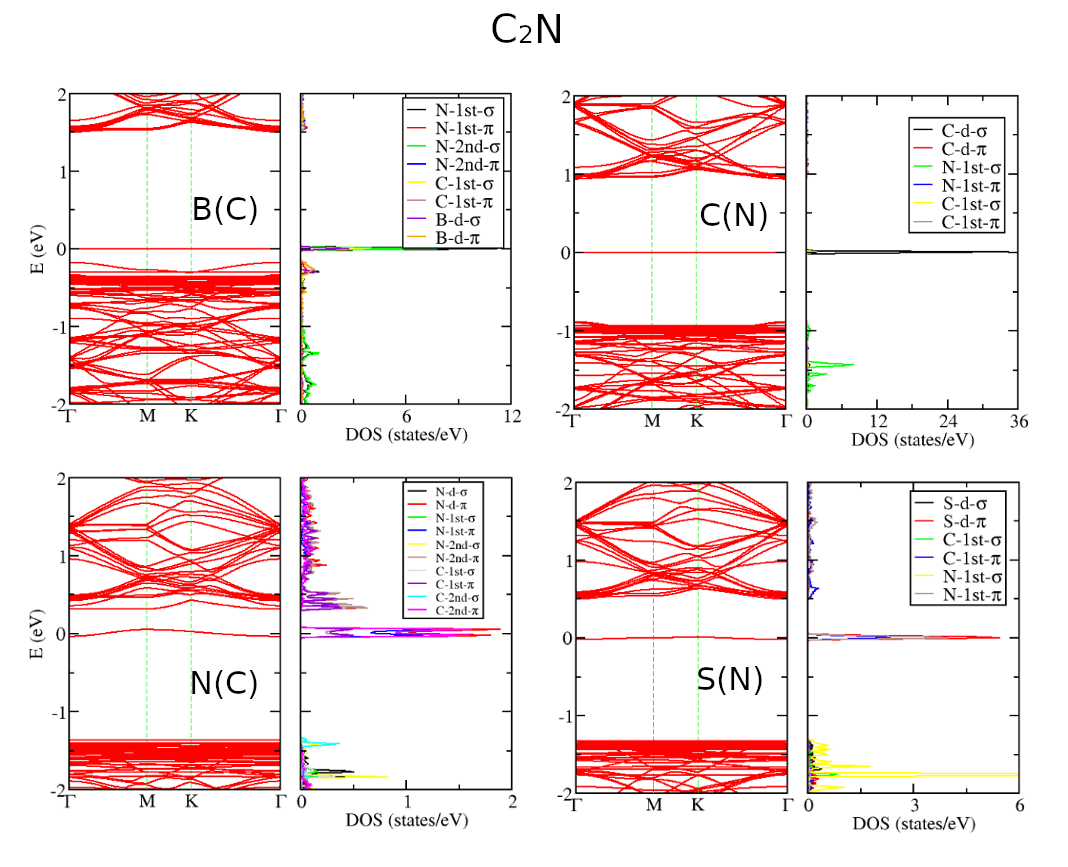}
    \caption{\label{fig:bands-unp-c2n} Band structure and projected density of states (PDOS) for different substitutional impurities in C$_2$N. (non-polarized calculations). See the text and Fig. \ref{fig:pure} for a description of the non-equivalent sites. The Fermi energy is set to zero in all cases and the PDOS color code is indicated in each panel. The labels `1st' and `2nd' label atoms which are first and second neighbors to the impurity, respectively. }
\end{figure}

\begin{figure*}
    \centering
    \includegraphics[width=0.9\textwidth]{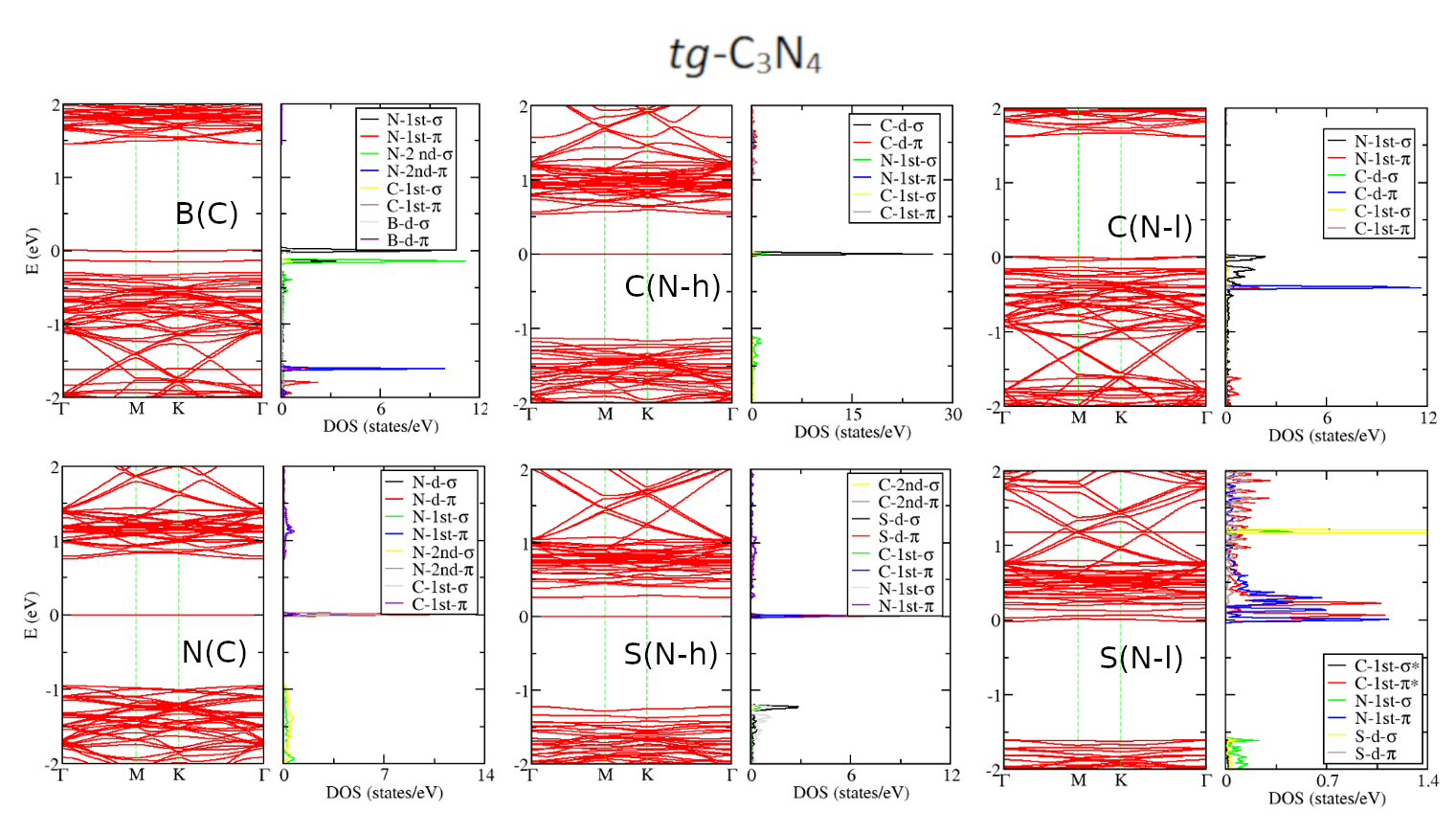}
    \caption{\label{fig:bands-unp-tg-c3n4} Band structure and projected density of states (PDOS) for different substitutional impurities in $tg$-C$_3$N$_4$. (non-polarized calculations). The PDOS labels follow the same notations as in Fig. \ref{fig:bands-unp-c2n}. }
\end{figure*}

\begin{figure*}
    \centering
    \includegraphics[width=0.9\textwidth]{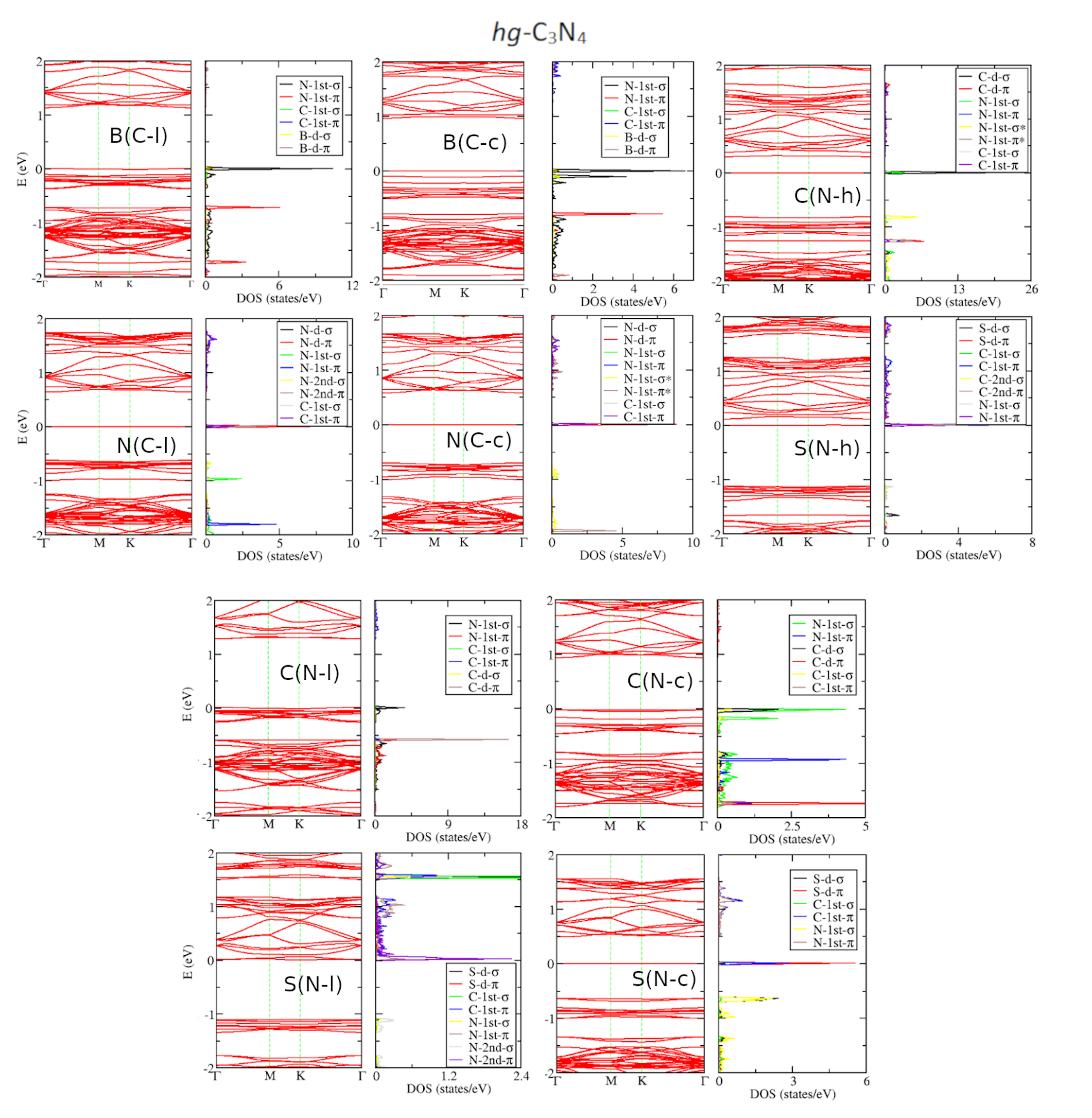}
    \caption{\label{fig:bands-unp-hg-c3n4} Band structure and projected density of states (PDOS) for different substitutional impurities in $hg$-C$_3$N$_4$. (non-polarized calculations). The PDOS labels follow the same rules as in Fig. \ref{fig:bands-unp-c2n}. }
\end{figure*}

We now discuss the electronic properties of the impurities. In order to facilitate the discussion, we begin with the results of non-polarized calculations and discuss the effects of spin polarization in the next sub-section. The band structures and projected density of states (PDOS) for each structure and impurity type are shown in Figs. \ref{fig:bands-unp-c2n}, \ref{fig:bands-unp-tg-c3n4} and \ref{fig:bands-unp-hg-c3n4}. As we can see, in all cases at least one impurity level can be found inside the band gaps. In a few cases, additional levels are found and for two cases in $hg$-C$_3$N$_4$, the ground-state energy level is nearly-degenerate.

We report the binding energies of each impurity level in Table \ref{tab:electronics}.
For donor (acceptor) impurities, they are calculated as the absolute value of the difference between the energy of the impurity level and the energy of the bottom of the conduction band (top of the valence band). It should be emphasized, however, that such a calculation for the binding energy based on Kohn-Sham energies is an approximation and more rigorous approaches consider total energy differences \cite{niquet-prb-2010}.
In the case of acceptors, we can see that the B(C) substitution can induce levels with moderate binding energies, between $0.09$ and $0.17$ eV, and the C(N) substitution can induce deep levels when the impurity is placed on a $h$ site, with binding energies as high as $1.13$ eV. When the C atom is placed on other N sites in the $tg$-, and $hg$-C$_3$N$_4$ structures, shallower levels are induced. These results can be understood in terms of the orbital composition of the valence bands of the pure materials, as discussed in sub-section \ref{subsec:pure} and shown in Fig. \ref{fig:pure}. From the point of view of effective mass theory \cite{kohn-luttinger,shindo-nara,gamble-prb,ullah2019exotic}, the wavefunction of an acceptor level should be a combination of Bloch states from the valence band of the pure system at $k$-points near its equivalent maxima (valleys). The PDOS plots in Fig. \ref{fig:pure} reveal that, in all structures, this band is strongly composed of $\sigma$ orbitals from N$_h$ sites, with smaller contributions from other N and C sites. Therefore, the effective potential induced by an acceptor impurity should be stronger when it is placed at a N$_h$ site, as the electronic density will be greater near it. This results in a higher binding energy for these cases. A similar behavior is seen in binary III-V materials such as $h$-BN, in which a strong sub-lattice sensitivity is observed for substitutional impurities \cite{ullah2019exotic}.

\begin{figure*}
    \centering
    \includegraphics[width=0.9\textwidth]{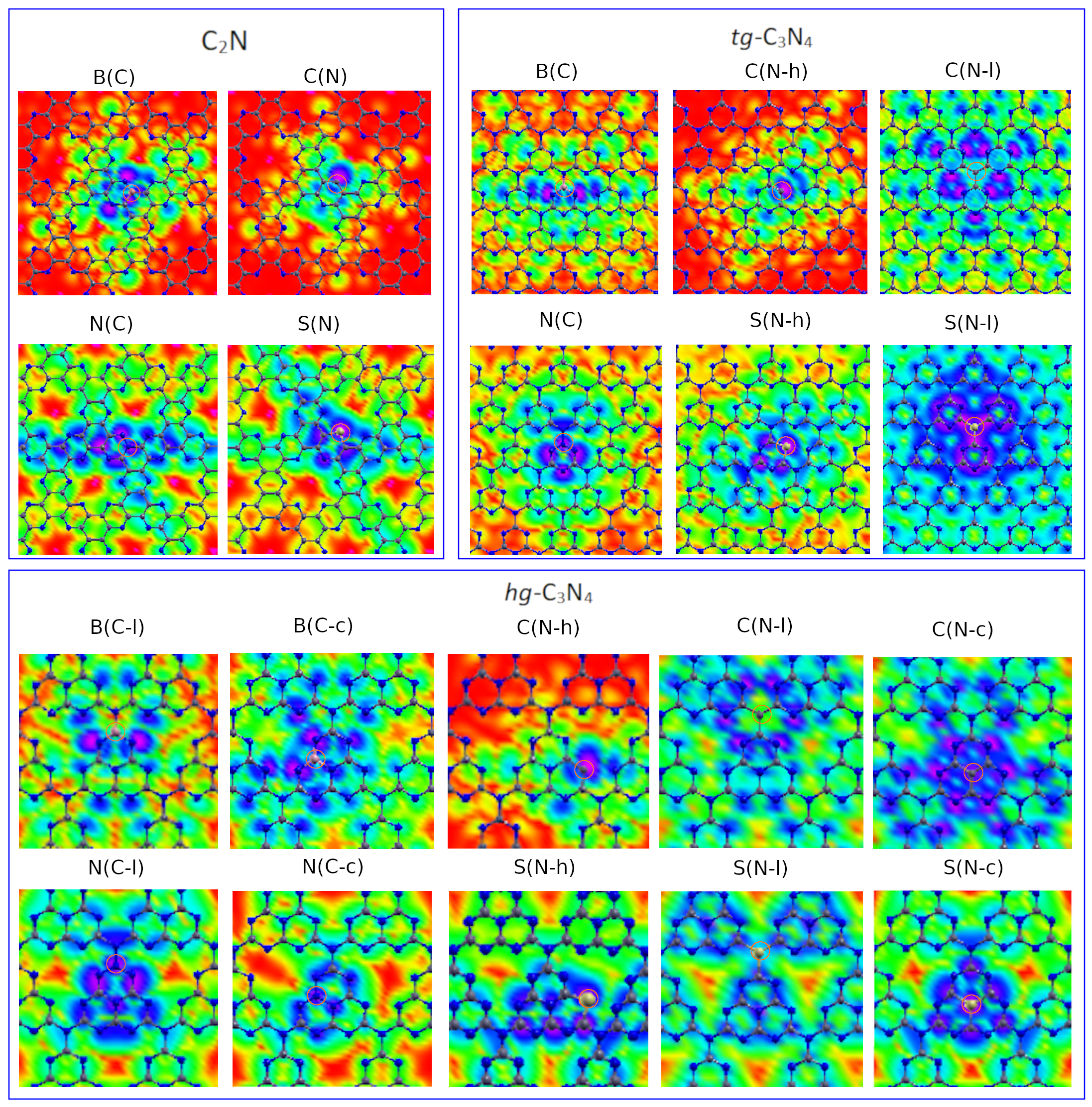}
    \caption{\label{fig:ldos} Local density of states (LDOS) for the ground state impurity level found in each case. When the ground state is nearly degenerate, the sum over the corresponding levels is shown. Gray, blue, pink and yellow spheres represent carbon, nitrogen, boron and sulfur atoms, respectively. In each case, the position of the impurity is indicated by an orange circle. The density is higher (lower) in violet (red) regions.}
\end{figure*}

Moreover, this result is supported by an analysis of the LDOS of the acceptor levels, shown in Figs. \ref{fig:bands-unp-c2n}-\ref{fig:bands-unp-hg-c3n4}, and the corresponding ground state wavefunctions shown in Fig. \ref{fig:ldos}. The defect-level wavefunctions (squared) were obtained from local density of states (LDOS) calculations with a narrow energy window spanning each level. In all cases, we can see that the impurity levels are composed mostly of $\sigma$ orbitals from either the impurity itself when it replaces a N$_h$ atom (in the case of C) or from surrounding N$_h$ atoms when it replaces a C atom (in the case of B). Note also that these wavefunctions are strongly directed towards the holes in the structures, as a consequence of their $\sigma$ character. This may be an indication that these acceptor impurities could modify the interaction between the material and adsorbates near the holes, which is a property that could further be explored for applications such as gas sensors \cite{novoselov2007graphene}.

\begin{table}[h]
    \setlength{\tabcolsep}{8.0pt}
    \setlength{\extrarowheight}{2.5pt}
    \centering
    \begin{tabular}{|l|c|c|}
        \hline
        Material                & Impurity & $E_b$ (eV) \\
        \hline
        \multirow{4}{*}{C$_2$N} & B(C)     & $0.17$    \\
                                & C(N)     & $0.88$    \\ \cline{2-3}
                                & N(C)     & $0.26$    \\
                                & S(N)     & $0.49$    \\
        \hline
        \multirow{6}{*}{$tg$-C$_3$N$_4$} & B(C)     & $0.29$, $0.16$    \\
                                       & C(N$_h$) & $1.13$             \\
                                       & C(N$_l$) & $0.14$, $0.09$    \\ \cline{2-3}
                                       & N(C)     & $0.75$             \\
                                       & S(N$_h$) & $0.26$             \\
                                       & S(N$_l$) & $0.11$  $0.06$    \\
        \hline
        \multirow{10}{*}{$hg$-C$_3$N$_4$} & B(C$_l$) & $0.09$            \\
                                        & B(C$_c$) & $0.30$, $0.19$, $0.07$ \\
                                        & C(N$_h$) & $0.90$, $0.07$   \\
                                        & C(N$_l$) & $0.04$            \\
                                        & C(N$_c$) & $0.22$ ($\times2$), $0.08$   \\ \cline{2-3}
                                        & N(C$_l$) & $0.64$            \\
                                        & N(C$_c$) & $0.58$            \\
                                        & S(N$_h$) & $0.11$            \\
                                        & S(N$_l$) & $0.03$ ($\times 2$)     \\
                                        & S(N$_c$) & $0.49$            \\
        \hline
    \end{tabular}
    \caption{\label{tab:electronics} Binding energies ($E_b$) of the impurity levels found inside the gaps in each case (non-magnetic calculations). Double (near) degenerate levels are represented by a ``$\times2$" label.}
\end{table}

In the case of donor impurities, the picture is a bit more complex. From Table \ref{tab:electronics}, we can see that both N(C) and S(N) impurities can induce deep levels. The binding energies from N(C) range from $0.26$ to $0.75$ eV, while those from S(N) range from $0.03$ to $0.49$ eV. Additionally, we can see that, for S(N), the $h$ sites lead to deeper levels in the C$_2$N and $tg$-C$_3$N$_4$ structures, while the same happens for the $c$ site in the $hg$-C$_3$N$_4$ structure. In contrast, in the latter structure, the N(C) impurity does not show a strong site sensitivity. Despite this complexity, these properties can also be understood in terms of the electronic structure of the pure materials, but in this case the relevant bands are the conduction bands. Again, from an effective mass point of view, the donor level wavefunctions should be combinations of Bloch states with k-points near the equivalent minima of these bands, which are also referred to as valleys. Note, however, that the orbital compositions of these bands are drastically different from those of the valence bands, as we have discussed above. Not only they are composed of out-of-plane $\pi$ orbitals instead of in-plane $\sigma$ orbitals, but the proportions between C and N atomic orbitals are different. In the C$_2$N and $tg$-C$_3$N$_4$ structures, the dominant contributions come from C and N$_h$ atoms in very similar proportions (per atom), while in the $hg$-C$_3$N$_4$ structures, they come from both non-equivalent C atoms and the N$_c$ atoms. Therefore, stronger binding energies are expected when the impurity is placed in these sites, as the electron feels a stronger impurity potential. As in the case of acceptors, these results are supported by an analysis of the PDOS of the impurity levels and the corresponding ground state wavefunctions (see Figs. \ref{fig:bands-unp-c2n}-\ref{fig:ldos}). These plots reveal that the impurity levels are composed mostly of $\pi$ orbitals from either the impurity itself, when the replaced atom had a significant contribution to the conduction bands, or from neighboring atoms when otherwise.

Next, we address the excited impurity levels and the near-degenerate ground states found in a few cases for both acceptor and donor impurities in the $tg$-, and $hg$-C$_3$N$_4$ structures. In the case of the $tg$-C$_3$N$_4$ structure, we can see from Fig. \ref{fig:pure} that both valence and conduction bands are actually two-fold degenerate at the $\Gamma$ point, which corresponds to their extrema. Therefore, impurity levels can be induced from both degenerate levels, with wavefunctions that are combinations of Bloch states from both of them and different binding energies according to the effective masses of the bands in the degenerate pair. Mixing between the bands is induced by the impurity potential. In the cases where the impurity potential is strong, such as C(N$_h$) and N(C) (as discussed above), we see a single impurity level inside the gap as a result of a strong band mixing. In contrast, in the case of B(C), C(N$_l$) and S(N$_l$), the potential is weaker as a result of a smaller electronic density near the impurity. This may explain the observation of an excited state inside the gap in these cases, with binding energies that are related to the different effective masses of the bands in the degenerate pair. We recall once again that a similar behavior is seen in $h$-BN, where its sub-lattice sensitivity to impurity placement leads to differences in the strength of the intervalley interaction and the resulting impurity level structure found inside the gap \cite{ullah2019exotic}. However, the multivalley nature of $h$-BN arises from the degeneracy of the valence and conduction bands at the $K$ and $K'$ points, while in our case it arises from two different bands that are degenerate at the same point ($\Gamma$). Finally, it is important to mention that S(N$_h$) could have a second impurity level inside the gap, as suggested in Fig. \ref{fig:bands-unp-tg-c3n4}. However, its PDOS analysis show small amplitudes near the impurity and its first neighbors, such that this state looks more compatible with the conduction band. 

For the $hg$-C$_3$N$_4$ structure, the situation is very different. We can see from Fig. \ref{fig:pure} that the valence and conduction bands are non-degenerate, but the conduction band minimum lies at the $K$ point, so it should also display multi-valley behavior. This may explain the near degenerate ground state found in S(N$_l$) as a result of weak inter-valley interaction. In the case of C(N$_h$), the excited state has a very different orbital composition from that of the ground state and the LDOS suggests it is an extended state. Finally, the B(C$_c$) and C(N$_c$) cases display the most complex level structures, with three levels inside the gap (two of them are nearly degenerate in C(N$_c$)). Given the low-bandwidth of the valence band of the pure material in this case, the multi-valley structure near the $M$ point could be relevant. As such, the contributions from the three non-equivalent $M$ points could lead to the three observed levels, with reduced binding energies and low splittings as a result of weak inter-valley interaction.

Curiously, the pure C$_2$N structure has a low-dispersive valence band, as in $hg$-C$_3$N$_4$, and the valence and conduction bands have degeneracies at the $\Gamma$ point, as in $tg$-C$_3$N$_4$. Still, we only find a single impurity level inside the gap in all cases. As binding energies are moderate in this case, with values from $0.17$ to $0.88$ eV, we believe that the impurity potential is strong enough to place excited levels outside the gap, mixed within the bands. Therefore, calculations based on generalizations of multi-valley effective mass theory, which include different k-points and band degeneracies, may be able to provide a deeper insight in the fine structure of the impurity levels in all cases. Such calculations are beyond the scope of this work.

Finally, notice that the presence of impurity levels may induce modifications in the optical properties of these materials, as commonly found in other 2D and conventional materials \cite{katzir-prb, bourrellier-nanolett, makaremi2018band, zheng2015lanthanide, gruber-science-1997,davies-prb-1992,martin-apl-1999}. New peaks may be present in the absorption spectrum due to new optical transitions between the valence band and the impurity levels or between the impurity levels and the conduction band. Such transitions will lie below the optical gap of the pure systems and, given the wide range of binding energies reported in Table \ref{tab:electronics}, we can see that substitutional doping may be an interesting route for tuning the optical properties of these materials.

\subsection{\label{subsec:electronics} Spin Polarization and Magnetic Properties}

Now we discuss the results of the spin-polarized calculations and the resulting magnetic properties. In fact, all impurity configurations we have considered have a magnetic ground state, with a magnetic moment of $1.0 \ \mu_B$ per unit cell and total energies per unit cell that are more than $100$ meV larger than the corresponding non-magnetic configurations.


\begin{figure}
    \centering
    \includegraphics[width=0.5\textwidth]{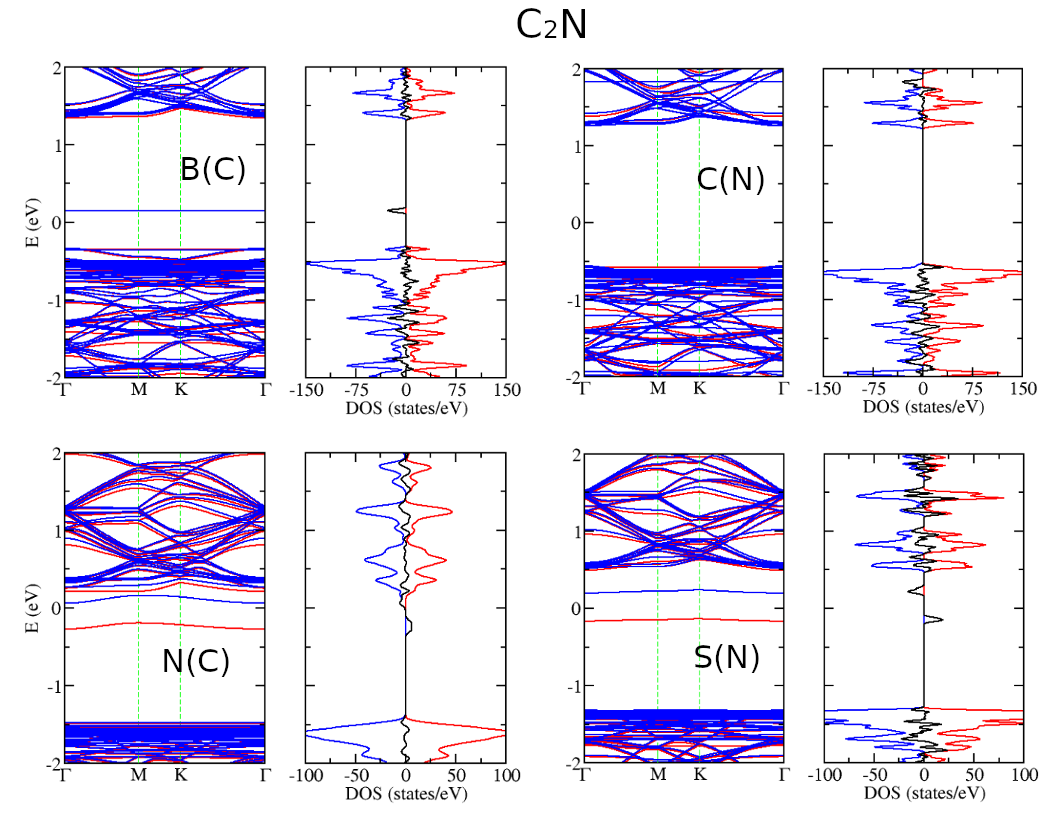}
    \caption{\label{fig:bands-c2n-pol} Band structure and total density of states (DOS) for each impurity case in the C$_2$N structure (spin-polarized calculations). Blue (red) lines correspond to spin up (spin down) states. In the DOS plots, the black curve corresponds to the difference between the spin up and down components.}
\end{figure}

\begin{figure*}
    \centering
    \includegraphics[width=\textwidth]{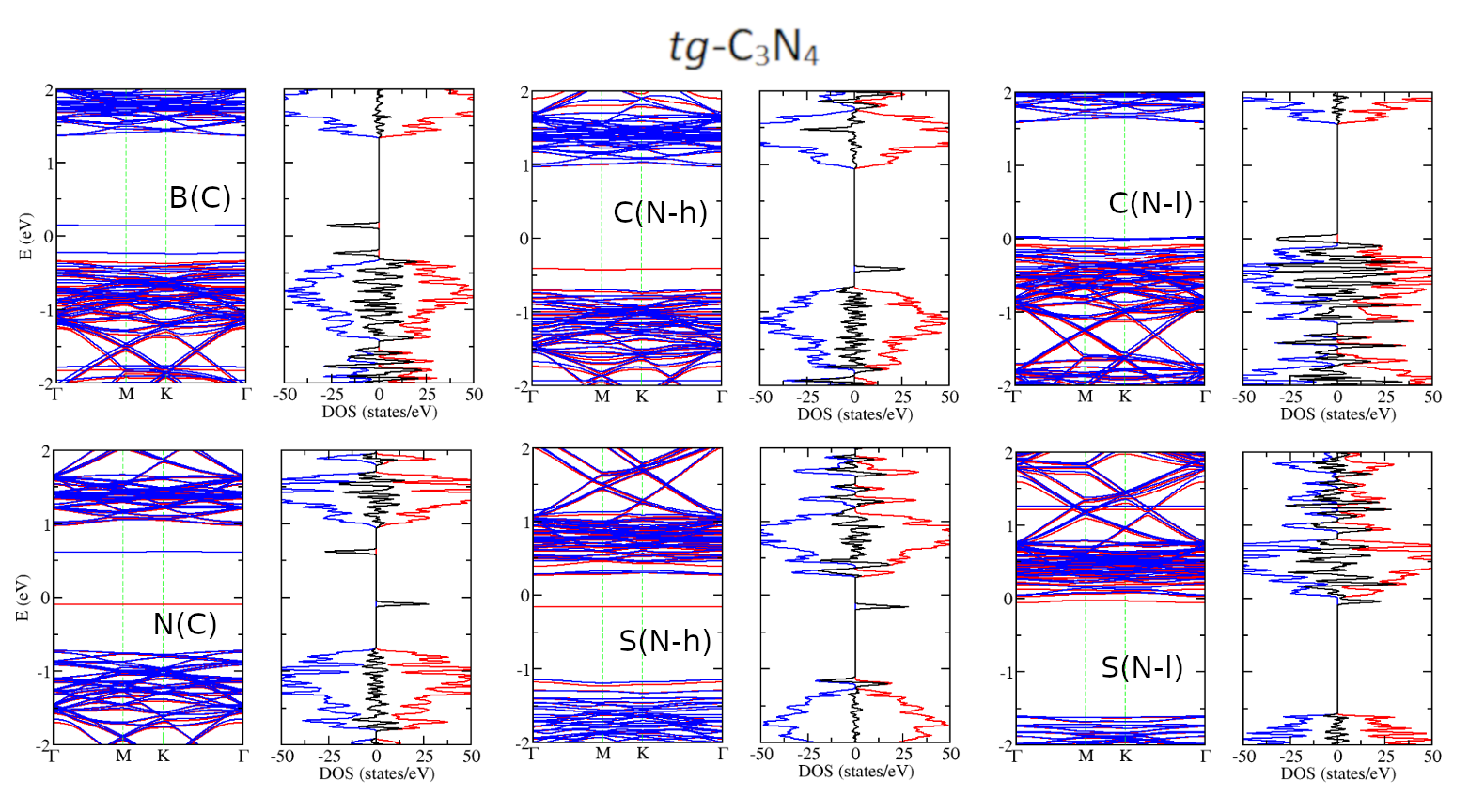}
    \caption{\label{fig:bands-tg-pol} Band structure and total density of states (DOS) for each impurity case in the $tg$-C$_3$N$_4$ structure (spin-polarized calculations). The color code is the same of Fig. \ref{fig:bands-c2n-pol}.}
\end{figure*}

\begin{figure*}
    \centering
    \includegraphics[width=\textwidth]{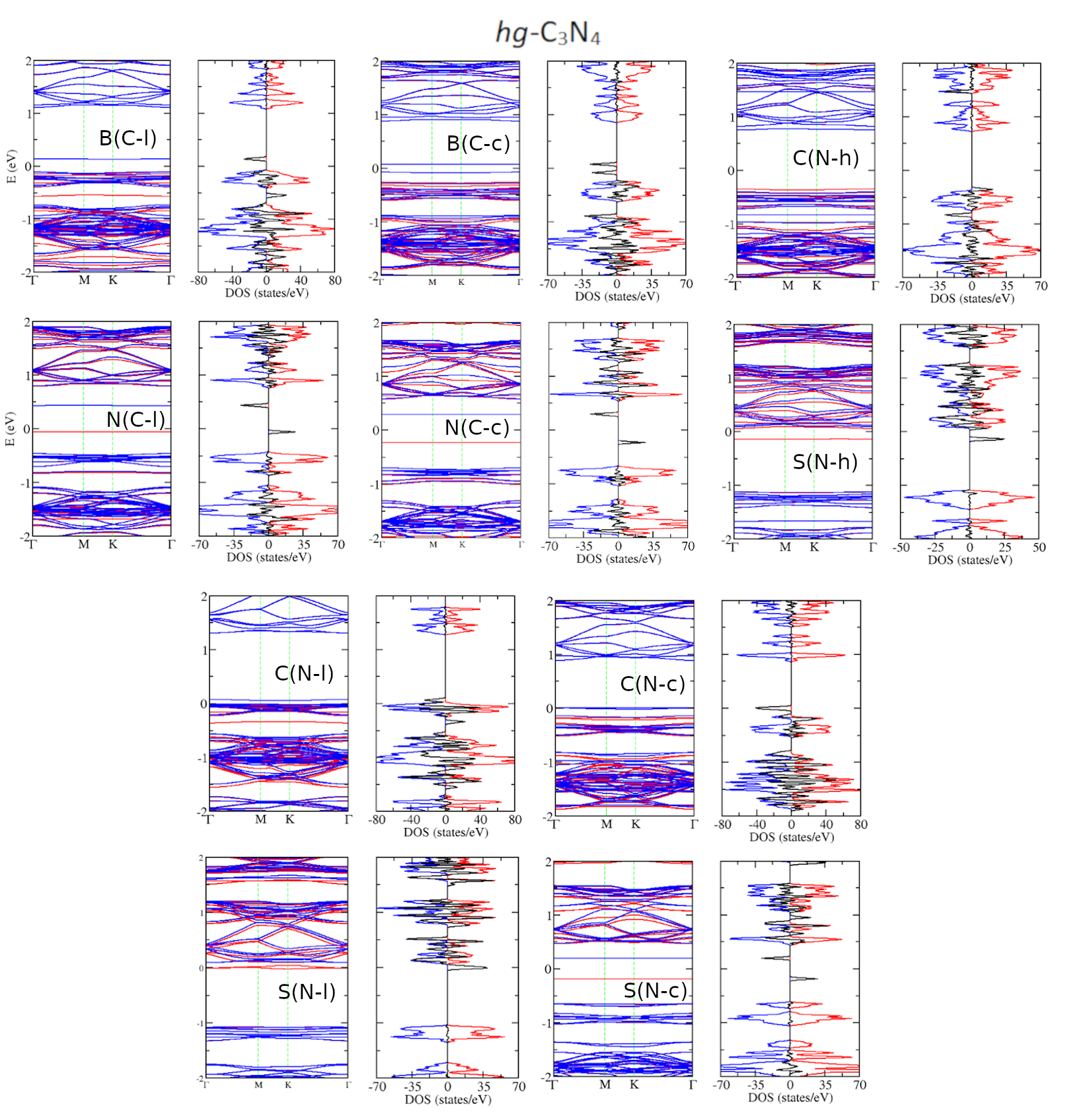}
    \caption{\label{fig:bands-hg-pol} Band structure and total density of states (DOS) for each impurity case in the $hg$-C$_3$N$_4$ structure (spin-polarized calculations). The color code is the same of Fig. \ref{fig:bands-c2n-pol}.}
\end{figure*}

The electronic band structures and total density of states (DOS) of the magnetic configurations are presented in Figs. \ref{fig:bands-c2n-pol}, \ref{fig:bands-tg-pol} and \ref{fig:bands-hg-pol}. As we can see, the most important effect of the polarization is a spin-splitting of the impurity levels, such that the magnetization comes from an imbalance between the occupations of the spin up and down levels. This is expected, since these levels are non-dispersive and they either lie at the Fermi energy or very close to it in the non-polarized calculations. As such, the density of states at the Fermi energy is large and a magnetic ground state is favored. Moreover, we can see that the magnitude of the splitting is different in each case and, in a few ones, it is so strong that the levels are actually removed from inside the band gap. In fact, this magnitude is related to the binding energy of each level, as electrons in deeper levels are more localized near the impurity and feel a stronger Coulomb interaction.

To exemplify these general features, consider the case of C$_2$N, for which the unpolarized and polarized band structures are shown in Figs. \ref{fig:bands-unp-c2n} and \ref{fig:bands-c2n-pol}, respectively. The C(N) impurity state displays the largest binding energy in the unpolarized calculation, which results in the strongest spin splitting in the polarized calculation, and the levels are actually taken away from the gap. This splitting is much weaker for B(C) and the donor impurities (N(C) and S(N)), where the levels remain (mostly) inside the gap and the imbalance between spin up and down occupations can clearly be seen. These properties can also be understood in terms of the impurity wavefunctions shown in Fig. \ref{fig:ldos}. Note that the wavefunction of the C(N) impurity state is more localized than those of the other three impurities. As such, the effects of the Coulomb interaction are stronger in the former case, resulting in a larger spin splitting. Naturally, similar comparisons can be drawn for impurities in the tg-, and hg-C$_3$N$_4$ structures.

Additionally, these impurities could further be investigated as sources of $\sigma$ or $\pi$ local magnetism, depending on their behavior as acceptors or donors, respectively. This is a topic of particular interest, as current technologies are mostly based on heavy metal atoms and $d$ or $f$ orbitals. In fact, the combination of doping with different impurities and the use of different sublattices could allow the tuning of the magnetic coupling between them. The effects of disorder and the defect concentration should also be investigated, in a similar fashion to the behavior of point defects in graphene systems \cite{nair2012spin,nair2013dual}. To model such disordered systems, \textit{ab-initio} methods become prohibitive and other approaches are needed, such as semi-empirical tight-binding calculations that are beyond the scope of this work.

Finally, note that two defect configurations display two non-degenerate levels for the same spin channel inside the gap: B(C) in $tg$-C$_3$N$_4$ and B(C-c) in $hg$-C$_3$N$_4$. Such configurations could potentially be explored as two-level systems for single photon emission, in a similar fashion to recent proposals based on defect complexes in TMDCs and $h$-BN \cite{gupta2018two, he2015single,koperski2015single,chakraborty2015voltage,bourrellier-nanolett,tran2016quantum}.


\section {\label{sec:conclusions} Conclusions}
In summary, we have employed DFT calculations to study the electronic and structural properties of substitutional impurities in the nanoporous materials C$_2$N, $tg$-, and $hg$-C$_3$N$_4$. Four types of impurities were considered: boron substitution on carbon sites (B(C)), carbon substitution on nitrogen sites (C(N)), nitrogen substitution on carbon sites (N(C)) and sulfur substitution on nitrogen sites (S(N)). We have found that the C(N) and B(N) substitutions are the most energetically favorable and induce small bond modifications in the vicinity of the impurity, while the S(N) induces strong lattice distortions. 

All impurities induce defect levels inside the band gap of these materials and their electronic properties are very different depending on the behavior of the impurity as an acceptor (B(C) and C(N)) or a donor (N(C) and S(N)). PDOS calculations indicate that acceptor (donor) impurities are composed only of $\sigma$ ($\pi$) orbitals from the impurity itself and/or neighboring sites, closely following the orbital compositon of the valence (conduction) band of the pure materials. For that reason, acceptor wavefunctions have in-plane amplitudes and are strongly directed towards the structural holes (pores) of these materials, while donor wavefunctions have out-of-plane amplitudes and are spread throughout neighboring atoms, as confirmed by our LDOS calculations. These properties could potentially be explored to modify the interaction between the material and adsorbates near the holes for applications such as gas sensors.

Moreover, the impurities display a strong site sensitivity, in a similar fashion to the sublattice sensitivity found in $h$-BN \cite{ullah2019exotic}. Depending on the atom replaced by the impurity and its orbital contribution to the valence or conduction band of the pure material, the induced impurity levels can either be shallow or deep and ground state binding energies range from $0.03$ to $1.13$ eV in non-magnetic calculations. Additionally, new optical transitions may arise between the valence band and the impurity levels or between the impurity levels and the conduction band, so we see that substitutional impurities offer an interesting route for tuning the optical properties of these materials, as in color centers in diamond and other defects in conventional materials \cite{gruber-science-1997,davies-prb-1992,martin-apl-1999}. 

Finally, spin-polarized calculations reveal that all impurity configurations have a magnetic ground state with a total moment of $1.0 \ \mu_B$ per unit supercell, which appears due to a spin splitting of the impurity levels and an imbalance between their occupations. The magnitude of the splitting also displays a strong site sensitivity, as electrons in deeper levels are more localized and feel a stronger Coulomb interaction. Curiously, in a few configurations, more than one impurity level can be found inside the gap. Two of them (B(C) in $tg$ and B(C$_c$) in $hg$-C$_3$N$_4$) have two non-degenerate levels of the same spin channel and could potentially be explored as two-level systems for single photon emission, following similar proposals recently made on defect complexes on TMDCs and $h$-BN.

\clearpage

\begin{acknowledgements}
We thank CNPq, CAPES, FAPEMIG, FINEP, FAPERJ, INCT Carbon Nanomaterials, PEDECIBA Quimica, CSIC, and
ANII Uruguayan institutions for financial support. SU wants to dedicate this work to Prof. Pervez Hoodbhoy for his relentless dedication to science.  
\end{acknowledgements}

\nocite{*}

\bibliography{bibliography}

\end{document}